\documentclass[useAMS,usenatbib,usegraphicx]{mn2e}

\newcommand{\chandra}{\textit{Chandra}}
\newcommand{\rrata}{RRAT~J1819$-$1458}
\newcommand{\rratb}{RRAT~J1846$-$0257}
\newcommand{\rratc}{RRAT~J0847$-$4316}
\newcommand{\RRATb}{J1846$-$0257}
\newcommand{\RRATc}{J0847$-$4316}
\newcommand{\RRATa}{J1819$-$1458}

\newcommand{\expnt}[2]{\ensuremath{#1 \times 10^{#2}}}   
\newcommand{\percmsq}{\ensuremath{{\rm cm}^{-2}}}
\newcommand{\ergsec}{\ensuremath{{\rm erg\,s}^{-1}}}
\newcommand{\NH}{\ensuremath{N_{\rm H}}}
\def\mc {\multicolumn}
\newcommand{\phn}{\phantom{0}}

\newcommand{\apjs}{ApJS}
\newcommand{\apj}{ApJ}
\newcommand{\apjl}{ApJ}
\newcommand{\nat}{Nature}
\newcommand{\mnras}{MNRAS}

\newcommand{\aap}{A\&A}

\newcommand{\procspie}{Proc.\ SPIE}
\newcommand{\apss}{Ap\&SS}
\newcommand{\araa}{ARAA}

\title{Upper Limits on X-ray Emission from Two Rotating Radio
  Transients}

\author[D.~L.~Kaplan et al.]{D.~L.~Kaplan$^{1,2}$\thanks{Email:
    dkaplan@kitp.ucsb.edu}, P.~Esposito$^{3,4}$, S.~Chatterjee$^{5}$,
  A.~Possenti$^{6}$, M.~A.~McLaughlin$^{7,8,9}$,\newauthor
F.~Camilo$^{10}$,
    D.~Chakrabarty$^{11}$, and P.~O.~Slane$^{12}$\\
$^{1}$ KITP, Kohn Hall, University of  California, Santa Barbara, CA
  93106, USA.\\
$^{2}$ Hubble Fellow\\
$^{3}$  INAF/IASF-Milano, Via E. Bassini 15, I-20133 Milano, Italy\\
$^{4}$  INFN-Pavia, Via Bassi 6, I-27100 Pavia, Italy\\
$^{5}$  Department of Astronomy, Cornell University, Ithaca, NY 14853, USA.\\
$^{6}$  INAF-Osservatorio Astronomico di Cagliari, localit{\`a} Poggio dei Pini, strada 54, I-09012 Capoterra, Italy\\
$^{7}$ Department of Physics, West Virginia Univerisity, Morgantown,
  WV 26506, USA.\\
$^{8}$ National Radio Astronomy Observatory, Green Bank, WV 24944,
  USA.\\
$^{9}$ Alfred P. Sloan Research Fellow\\
$^{10}$ Columbia Astrophysics Laboratory, Columbia University, New York, NY 10027, USA\\ 
$^{11}$ Kavli Institute for Astrophysics and Space Research and Department of Physics,
  Massachusetts Institute of Technology, Cambridge, MA 02139, USA.\\
$^{12}$ Harvard-Smithsonian Center for Astrophysics, Cambridge, MA 02138, USA.\\
}

\begin{document}

\date{MNRAS, in press}

\maketitle

\begin{abstract}
X-ray emission from the enigmatic Rotating RAdio Transients (RRATs)
offers a vital clue to understanding these objects and how they relate
to the greater neutron star population. An X-ray counterpart to
\rrata\ is known, and its properties are similar to those of other
middle-aged (0.1\,Myr) neutron stars.  We have searched for X-ray
emission with \chandra/ACIS at the positions of two RRATs with
arcsecond (or better) localisation, \RRATc\ and \RRATb.  Despite deep
searches (especially for \rratb) we did not detect any emission with
0.3--8\,keV count-rate limits of $1\,{\rm counts\,ks}^{-1}$ and
$0.068\,{\rm counts\,ks}^{-1}$, respectively, at 3$\sigma$ confidence.
Assuming thermal emission similar to that seen from \rrata\ (a
blackbody with radius $\approx 20\,$km), we derive effective
temperature limits of 77\,eV and 91\,eV for the nominal values of the
distances and column densities to both sources, although both of those
quantities are highly uncertain and correlated. If we instead fix the
temperature of the emission (a blackbody with $kT=0.14\,$keV), we
derive unabsorbed luminosity limits in the 0.3--8\,keV range of
$\expnt{1}{32}\,\ergsec$ and $\expnt{3}{32}\,\ergsec$.  These limits
are considerably below the luminosity of \rrata\
($\expnt{4}{33}\,\ergsec$), suggesting that RRATs~\RRATc\ and \RRATb\
have cooled beyond the point of visibility (plausible given the
differences in characteristic age).  However, as we have not detected
X-ray emission, it may also be that the emission from RRATs~\RRATc\
and \RRATb\ has a different character from that of \rrata.  The two
non-detections may prove a counterpoint to \rrata, but more detections
are certainly needed before we can begin to derive general X-ray
emission properties for the RRAT populations.
\end{abstract}

\begin{keywords}
pulsars --
stars: neutron --
X-rays: stars 
\end{keywords}

\section{Introduction}
The radio-transient sky is one of the least explored frontiers in
astronomy.  There have been many searches for transient signals, but
few astrophysical sources of radio bursts have been confirmed
(\citealt*{clm04}; \citealt{hlk+05}). However, \citet{mll+06} reported the discovery
of eleven objects characterised by the recurring although
unpredictable emission of single, dispersed radio pulses (additional
objects have been discovered by \citealt{dcm+09} and others).  The
unique dispersion measures identified them as astronomical signals,
and the peak flux densities (ranging from 100\,mJy to 3.6\,Jy) made
these transients among the brightest radio sources in the universe.
The most natural interpretation of these sources is that they are
rotating neutron stars (NSs) which emit radio bursts sporadically:
hence the name Rotating RAdio Transients (RRATs).

Based on their ephemeral nature, the total number of RRATs in the
Galaxy ($\sim 10^5$; \citealt{mll+06,kk08}) might exceed that of the
radio pulsars, thus representing a large fraction of the population of
neutron stars with age smaller than $\sim 10^7$ yr. That calls for a
deep investigation of these objects, aimed at understanding their
emission properties,  environments, and evolutionary links
(if any) with other classes of neutron stars.  It has been proposed
that RRATs could be an extreme manifestation of processes already
observed in at least some ordinary pulsars: e.g., giant pulses
\citep{kbm+06},  nulling \citep*{zgd07}, or the fading of the radio
signal of an old pulsar while approaching the pair-production death
line \citep{zgd07}.  A particularly interesting hypothesis is that
RRATs may be distant middle-aged (spin-down ages $1-5\times 10^5$\,yr)
ordinary pulsars (such as PSR~B0656+14), whose emission is only
detectable when particularly bright pulses occur \citep{wsrw06}.
According to other models, the RRATs do not belong to the population
of the rotation-powered pulsars: instead associations with the
steadily-emitting magnetars \citep[e.g.,][]{mll+06}, transient
magnetars \citep{mlk+09}, or thermally-emitting isolated neutron stars
(INS; \citealt*{ptp06}) were suggested.  Even more exotic hypotheses
explain the bursting behavior of the RRATs with the sporadic effects
of an asteroid or  radiation belt surrounding the pulsar
\citep[e.g.,][]{li06,cs08,lm07}.

The observation of the counterpart to a RRAT in the X-ray band could
be crucial for discriminating between many of the models reported
above, since they predict distinct X-ray spectral signatures for the
source.  Unfortunately, only \rrata\ (with spin period $P=4.26$\,s,
dipole magnetic field $B= 5\times 10^{13}$\,G, characteristic age $
\tau=1.2\times 10^5\,$ yr, and spin-down luminosity $\dot E= 3\times
10^{32}\,\ergsec$; for definitions of these parameters see
\citealt{lk04}) has been detected at high energies. It was
serendipitously found in \chandra\ observations of the (almost
certainly unrelated) supernova remnant G15.9+0.2. Additional long
pointed integrations revealed that \rrata\ has a thermal spectrum with
$kT=0.14$\,keV and an unabsorbed X-ray luminosity $L_X \approx
4\times10^{33}\,\ergsec$ \citep{mrg+07}, which is larger than the
spin-down luminosity. The spectrum differs from that of the magnetars,
which are hotter \citep[$kT \sim 0.3-0.6$\,keV][]{wt06}, although it
may be similar (at times) to that of the transient magnetar
XTE~J1810$-$197 ($kT \sim 0.15-0.18$\,keV;
\citealt{ims+04,gh05}). Instead, the spectrum of \rrata\ is comparable
in temperature \citep{kpz+01} and X-ray luminosity to the
similarly-aged radio pulsar B0656+14 (which itself is consistent with
being a nearby RRAT; see above). It is also similar to the slightly
older radio-quiet INS ($kT=0.05-0.1$~keV at ages of $\approx 0.5$~Myr;
see \citealt{haberl07,kaplan08}). In addition, the spectrum displays a
broad absorption line \citep{mrg+07} resembling the lines detected in
many of the INS \citep{haberl07,vkk07}.

The case of \rrata\ illustrates the potentiality of X-ray observations
in unveiling possible connections between RRATs and other NS
populations.  However, until recently these studies have been
significantly limited by the lack of precise positions for most of the
sources.  \rrata\ was one of the only three RRATs whose celestial
coordinates had been precisely determined through pulse timing (e.g.,
as discussed in \citealt{lk04}). All the other RRATs were only
localised to within the primary beam  of the Parkes telescope:
$\approx 14\,$arcmin diameter for the observing frequency of 1.4\,GHz.  That
prevented follow-up multi-wavelength observations and, in turn, the
possibility of constraining the spectrum of the RRATs and their
primary energy stores.

Dedicated timing observations presented in \citet{mlk+09} have more
than doubled the sub-sample of the RRATs with phase-connected timing
solutions and hence high precision rotational and astrometric
parameters\footnote{These timing solutions were derived by fitting a
timing model to pulse times-of-arrival using {\sc TEMPO}, as is done
for normal radio pulsars, but using single pulses rather than average
profiles. To avoid underestimating the positional uncertainties, the
quoted 1-$\sigma$ positional uncertainties have been calculated by
requiring the reduced $\chi^2$ values of the residuals to be equal to
one (see \citealt{mlk+09} for more details). As a confirmation of the
accuracy of this technique, we note that the position of the X-ray
counterpart to \rrata\ (confirmed via the identical periodicity of its
X-ray pulsations) is well within the radio timing error circle
\citep{mll+06,rmg+09}.}.  Two among the RRATs with a new timing
solution (\RRATc\ and \RRATb) are located in the $P-\dot{P}$ diagram
in a region almost devoid of ordinary pulsars (see Section
\ref{sec:obs} for timing parameters) but close to the area occupied by
the INS, lending support to the hypothesis (\citealt{kk08};
\citealt{mlk+09}) that they could be transition objects between
ordinary pulsars and INS.

Inspecting the $P-\dot{P}$ diagram (see e.g., Figure~4 in
\citealt{mlk+09}), \RRATc\ and \RRATb\ also appear the closest RRATs
to the location of \RRATa, the only RRAT detected in X-ray so far (see
above).  Here we present searches for X-ray emission from these two
particularly promising sources, based on targeted and archival {\it
Chandra} observations.  In what follows, all luminosities are
corrected for interstellar absorption.

\setlength{\tabcolsep}{1mm}
\begin{table}
\caption{Summary of X-ray Observations\label{tab:obs}}
\begin{tabular}{l c c c c c}
\hline
\mc{1}{c}{RRAT} & \mc{1}{c}{ObsID} & \mc{1}{c}{Date} &  \mc{1}{c}{Detector}& \mc{1}{c}{Exp.} &\mc{1}{c}{Off-Axis Angle} \\
 & & & & \mc{1}{c}{(ks)} & \mc{1}{c}{(arcmin)} \\
\hline
\RRATc & 7626 & 2007-02-08       & ACIS-I & 10.7 & 1.1    \\
\RRATb & \phn748 & 2000-10-15 & ACIS-S & 37.8 & 1.1 \\
 & 6686 & 2006-06-07  & ACIS-S & 55.3 & 1.8     \\
&7337 & 2006-06-05      & ACIS-S & 17.8 & 1.8       \\
&7338 & 2006-06-09      & ACIS-S & 40.1 & 1.8       \\
&7339 & 2006-06-12      & ACIS-S & 45.1 & 1.8       \\
\hline
\end{tabular}\\
The first two observations were taken in the default Faint mode and
with a 3.2\,s frame time.  The remaining observations were taken in
the Very Faint mode, and with a 1.8\,s frame time to reduce the effect
of pileup on the probably unrelated PSR~J1846$-$0258; see \citet*{hcg03} and
\citet{nsgh08}.
\end{table}

\begin{figure*}
\centerline{\includegraphics[width=0.5\textwidth]{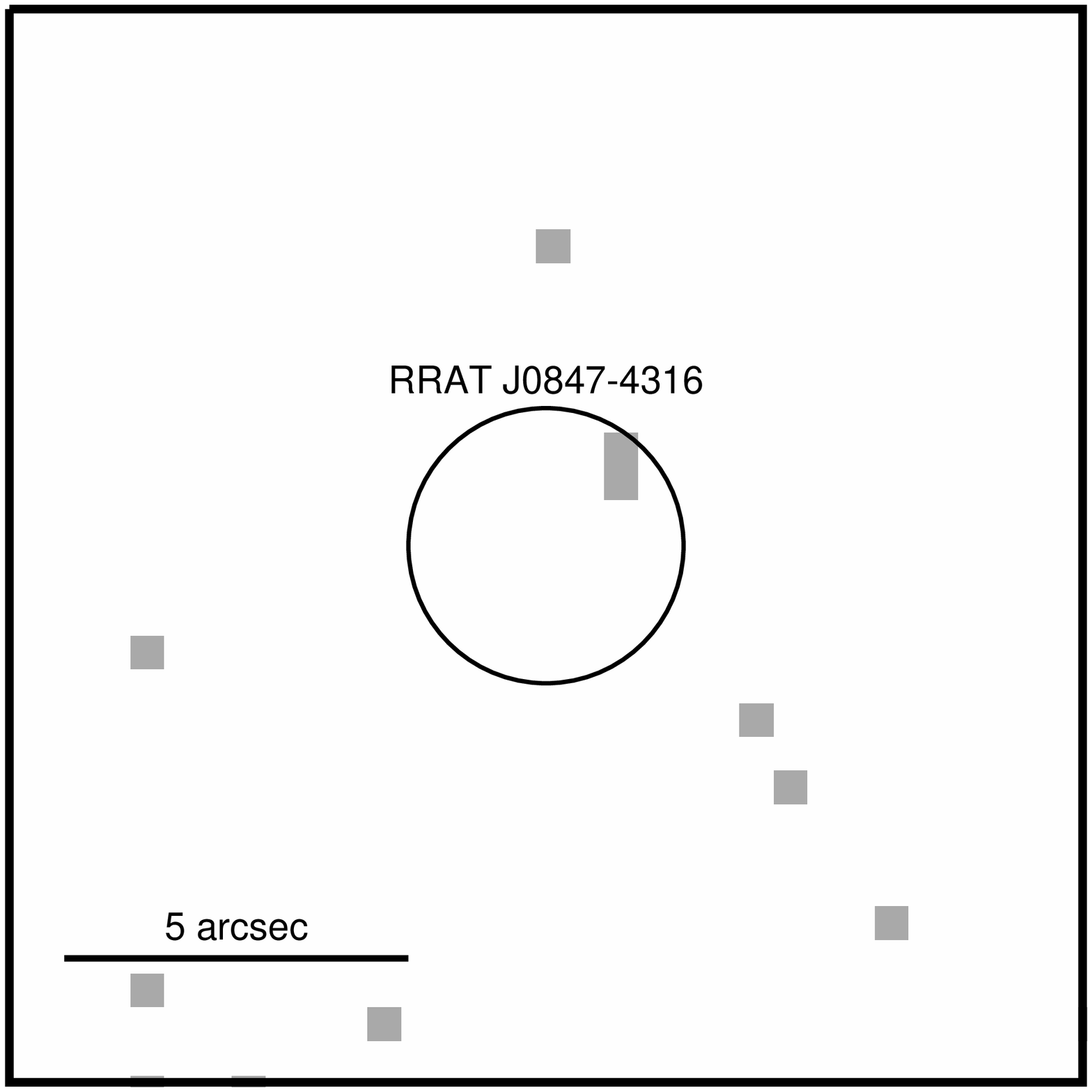}%
\includegraphics[width=0.5\textwidth]{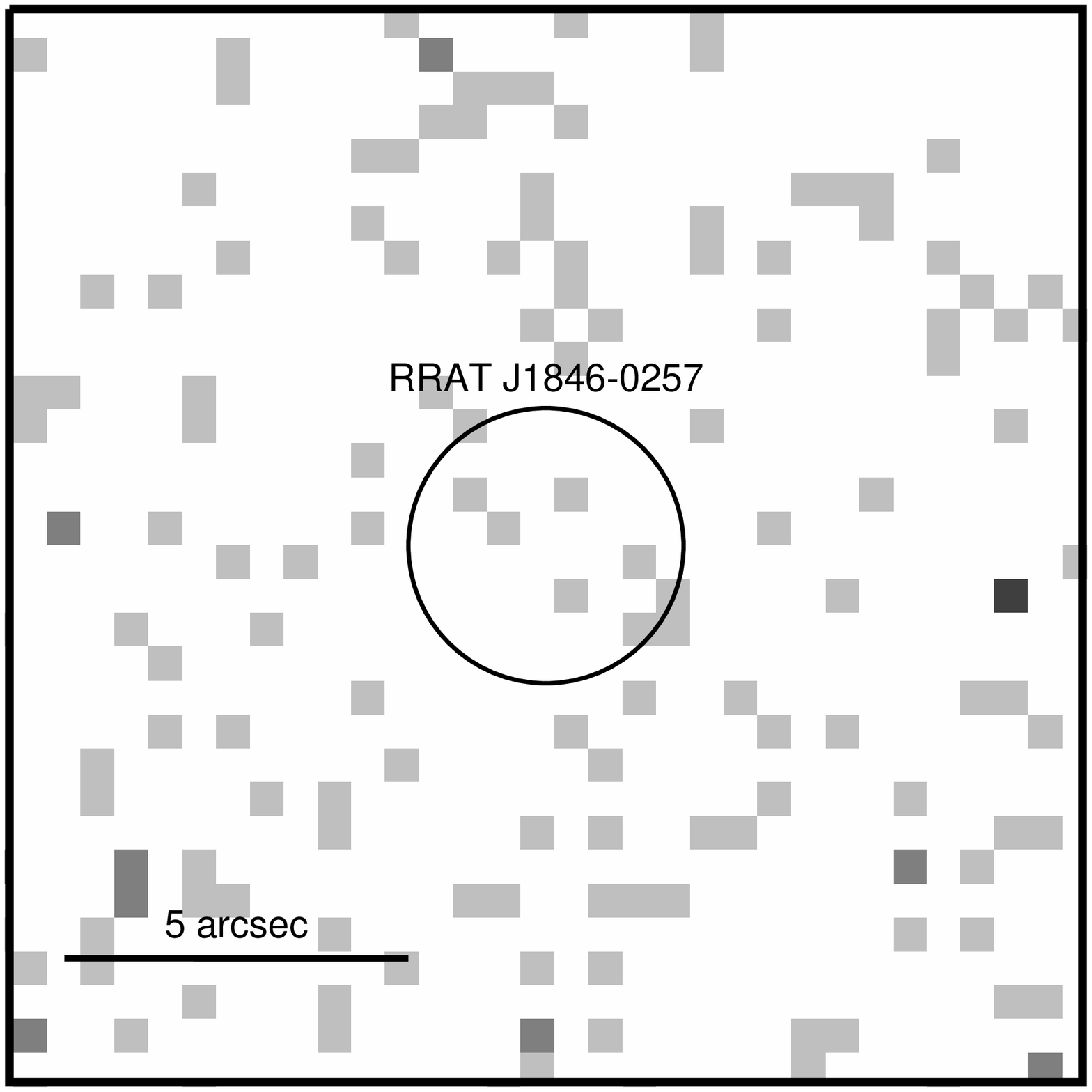}}
\caption{\chandra\ images of RRATs \RRATc\ and \RRATb\ in the
  0.3--8.0\,keV band.  \textit{Left:} 10.7\,ks \chandra/ACIS-I1 image
  of \rratc.  \textit{Right:} 158.3\,ks \chandra/ACIS-S3 image of
  \rratb, created from summing the observations from 2006 in
  Table~\ref{tab:obs}.  Overlaid on each is the region (2\,arcsec
  radius circle) selected for the analysis (see text). Both images are
  15\,arcsec square, with north up and east to the left. A 5\,arcsec
  scale bar is in the lower left of each image.
\label{fig:image}
}
\end{figure*}

\section{Observations and Upper Limits}
\label{sec:obs}
\subsection{\rratc}
We searched for an X-ray counterpart to \rratc\ ($P=5.978\,$s,
$B=\expnt{3}{13}\,$G, $\tau=0.8\,$Myr, and$\dot
E=\expnt{2}{31}\,\ergsec$) with the \textit{Chandra X-ray Observatory}
\citep{wtvso00}, as summarised in Table~\ref{tab:obs}.  As the
position was originally only known to $\pm7\,$arcmin, we used the
$16\,$arcmin$^2$ imaging array on the Advanced CCD Imaging
Spectrometer (ACIS-I; \citealt{gbf+03}).  The data were reduced in a
standard way using the {\sc CIAO} software package (version 4.1)
and the {\sc CALDB} 4.1 calibration files, restricting the energy
range to 0.3--8\,keV.  With the timing position of \rratc\ available
from \citet[][J2000: $08^{\rm h}47^{\rm m}57\fs33\pm0\fs05$,
$-43\degr16\arcmin56\farcs8\pm0\farcs7$]{mlk+09}, we were able to localise the
source to the ACIS-I1 detector.  The position is near a gap between
CCDs of the ACIS-I detector, but is not in the gap: examination of the
exposure map (which takes into account effective area maps of each
detector as well as the dither pattern of \chandra\ on the sky) shows
an effective area at the position of \rratc\ comparable to the average
of the ACIS-I1 CCD.  However, we see no source at the position of
\rratc, and searches using {\sc wavdetect} did not find any source
closer than $1.5\,$arcmin.  We must go out to a radius of $>1$\,arcsec
before there are any counts, and those counts appear consistent with
the background rate ($0.0300\pm0.0010\,{\rm counts\,arcsec}^{-2}$).
We compare this with the radio position uncertainty ($\approx
0.6\,$arcsec), the \chandra\ aspect uncertainty ($0.6\,$arcsec at 90 per
cent confidence, as this source is on-axis; there is no systematic aspect
offset listed for this observation), and the \chandra\ point-spread
function (90 per cent encircled energy radius of $\approx 1$\,arcsec at
1.5\,keV).  Using a generous extraction radius of $2\,$arcsec there are
2 counts with 0.4 expected from the background (see
Figure~\ref{fig:image}).  Given Poisson fluctuations
\citep{gehrels86}, we can place a 3$\sigma$ upper limit of $\approx
10$ counts coming from \rratc, or a count-rate limit of $\approx
1\,{\rm counts\,ks}^{-1}$.  Note that for \rrata\ about 10 per cent of the
X-ray flux is in an extended nebula going out to at least $13\,$arcsec
\citep{rmg+09}, but the contribution is small enough that our limits
are not affected by the size of our extraction region.

\subsection{\rratb}
The field of \rratb\ ($P=4.477\,$s, $B=\expnt{3}{13}\,$G,
$\tau=0.4\,$Myr, and $\dot E=\expnt{7}{31}\,\ergsec$) has been imaged
by various X-ray instruments during observations targeting the
probably unrelated (see Section~\ref{sec:disc}) X-ray pulsar
PSR~J1846$-$0258 and its supernova remnant Kes~75 (\citealt{ggg+08}
and references therein). Among the available data, those collected
with \chandra\ are the best suited to search for X-ray emission from
\rratb\ because of the superb angular resolution. Five observations
were carried out between 2000 and 2006 (see Table~\ref{tab:obs}) with
the source positioned on the back-illuminated ACIS-S3 chip. We again
reduced the data using {\sc CIAO}, applying aspect corrections and
filtering the data to reject time intervals of flaring background as
necessary; see \citet{hcg03} and \citet{nsgh08} for detailed
information about the datasets.  Figure~\ref{fig:image} shows the
image from the combined data. Source detection was performed with the
{\sc celldetect} and {\sc wavdetect} routines in {\sc CIAO}
over numerous energy bands on each individual dataset and on a
cumulative image that was obtained from the 2006
observations\footnote{Observation 748 was excluded because of its
different telemetry format and operating mode.}  with a total exposure
of 158\,ks. No X-ray source was found near the \rratb\ radio position
(J2000: $18^{\rm h}46^{\rm m}15\fs49\pm0\fs04$, $-02\degr57\arcmin36\farcs0\pm1\farcs8$). As with
\rratc\ we used a generous search radius of $2\,$arcsec, which combines
the radio position uncertainty, the \chandra\
position uncertainty and the \chandra\ point-spread function. Based on
the 8 events within this circle (background expectation is 7), this
upper limit is $\expnt{6.8}{-2}\,{\rm counts\,ks}^{-1}$ in the 0.3--8
keV energy band, at  3$\sigma$ confidence level (again based on
\citealt{gehrels86}).

\subsection{Luminosity Limits}
\label{sec:limits}
To interpret the count-rate limits derived in the previous sections,
we need a model for the X-ray spectrum of the RRATs, as well as the
distance and interstellar absorption along each line of sight.  For
\rratc\ we take the distance of 3.4$d_{3.4}$\,kpc based on the
observed dispersion measure and the \citet{cl02} electron density
model.  Using the 3-dimensional Galactic extinction model of
\citet*{dcllc03}, we find an optical extinction of $A_V\approx 3\,$mag.
This translates to a hydrogen column density of
$\NH=5\times10^{21}\,\percmsq$ (based on \citealt{ps95}).  This is
similar to the value of $\expnt{9}{21}\,\percmsq$ based on the
observed dispersion measure of $293\,{\rm pc\,cm}^{-3}$ and assuming
10 per cent ionisation (similar to the $\NH/DM$ ratio of \rrata), or the same
value from the \citet{dl90} {\sc H\,I} data-set (likely to be an upper
limit, since it is integrated through the Galaxy).  Overall, values of
$\expnt{(5-10)}{21}\,\percmsq$ seem likely, but the uncertainty is
large given the poorly known distance.

For \rratb, we use a distance of $5.2d_{5.2}\,$kpc estimated from the
dispersion measure.  The \citet{dcllc03} model gives an extinction of
$A_V\approx 9\,$mag which implies $\NH=1.6\times10^{22}\,\percmsq$.
We compare this with the DM-derived value of
$\expnt{0.7}{22}\,\percmsq$ or the {\sc H\,I}-derived value of
$\expnt{2}{22}\,\percmsq$.  We also have the value of
$\expnt{4}{22}\,\percmsq$ measured for PSR~J1846$-$0258/Kes~75, which
is likely at a comparable or slightly larger distance (5.1--7.5\,kpc;
\citealt{lt08}) as \rratb.  This gives us a wide range to consider,
$\expnt{(1-4)}{22}\,\percmsq$, again with the understanding that the
large uncertainty on the distance gives an additional contribution.

As an aside, we consider the possibility that PSR~J1846$-$0258 and
\rratb\ were members of a binary system that was disrupted by a
supernova.  At 5\,kpc their transverse separation of $2.5\,$arcmin is
3.6\,pc.  In this scenario, the explosion that produced \rratb\
0.4\,Myr ago ejected the progenitor of PSR~J1846$-$0258, which would
have been moving at a transverse velocity of $\sim 9\,{\rm
km\,s}^{-1}$ before the second supernova explosion that created the
Kes~75 remnant.  This velocity is actually very low for an object
ejected out of a binary, which typically move at several hundred
km\,s$^{-1}$ \citep[e.g.,][]{vcc04}, although we do not know the
radial velocity of either object.  Waiting $\sim 0.5\,$Myr for the
second supernova seems plausible, at least based on some studies of
binary evolution \citep{brown95}.  If PSR~J1846$-$0258 were actually
the second supernova of the system it could have interesting
implications, as the progenitors of magnetars are typically thought of
as quite massive \citep[e.g.,][]{khg+04,eml+04,gmgo+05,mcc+06} and
PSR~J1846$-$0258 shares some (but not all) characteristics of
magnetars \citep{ggg+08}, but being the second supernova would suggest
it is less massive than its companion.  Mass transfer or interactions
during binary evolution could likely resolve the discrepancy.

\begin{figure*}
\includegraphics[width=0.5\textwidth]{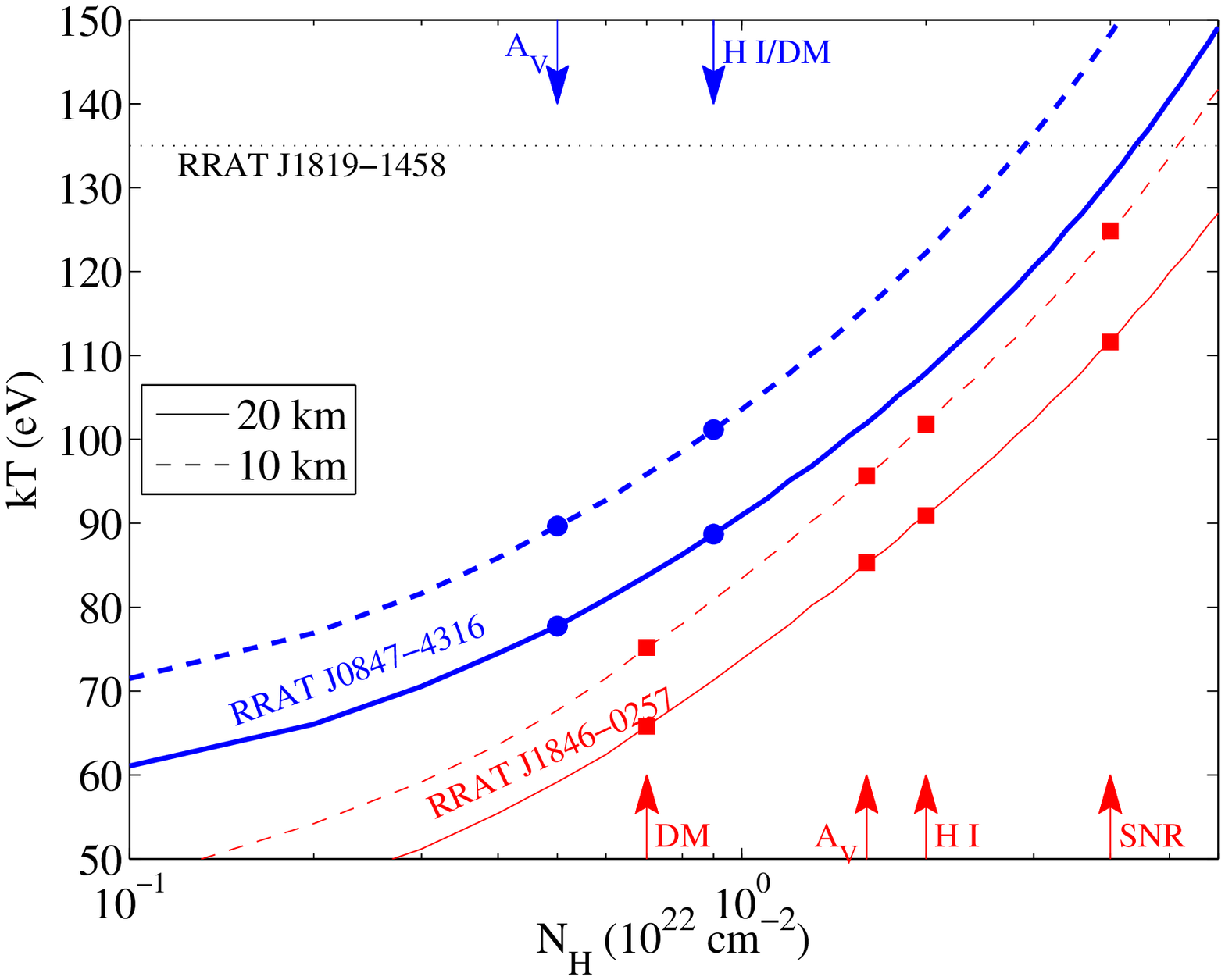}\includegraphics[width=0.5\textwidth]{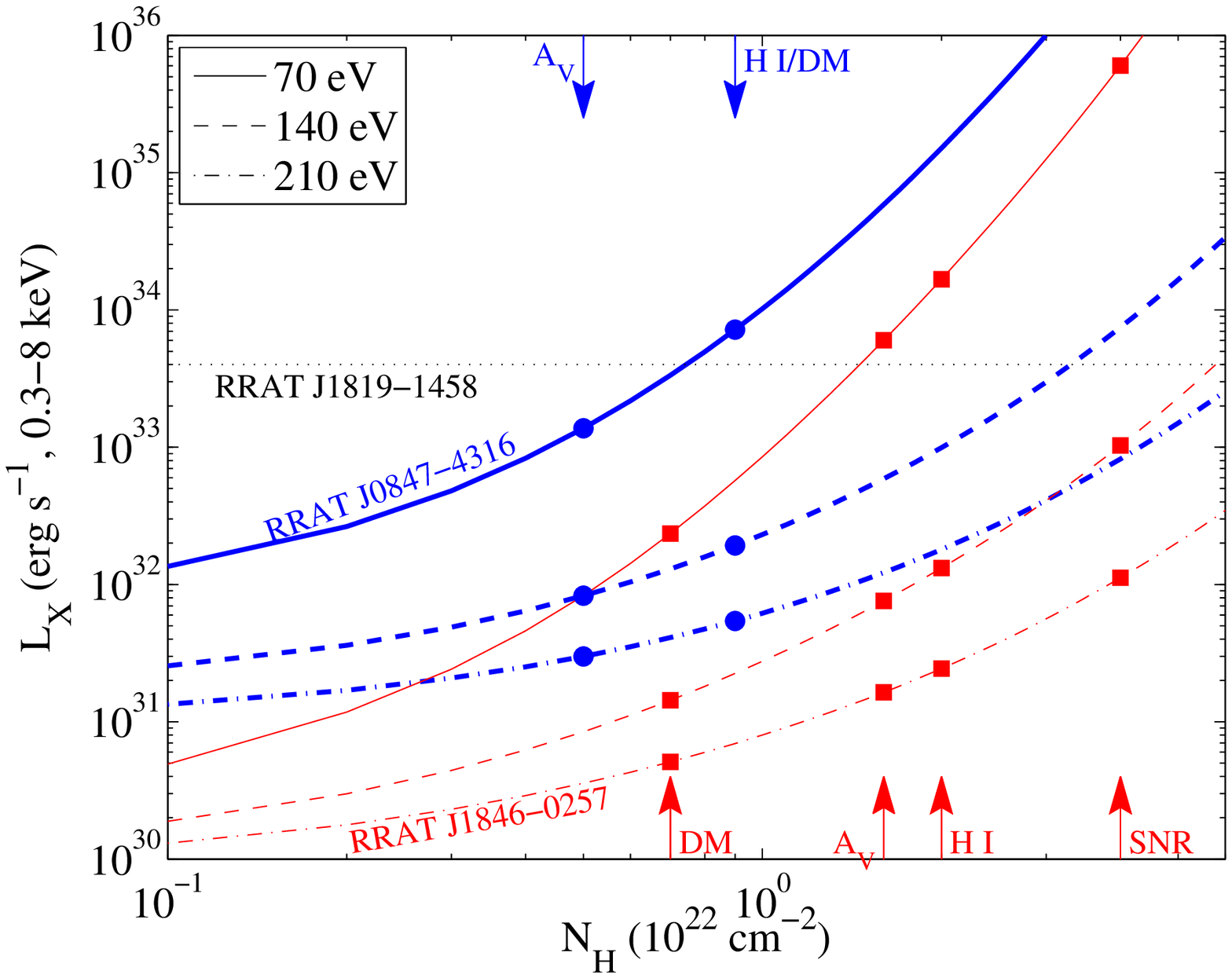}
\caption{\textit{Left:} Upper limits to the blackbody effective
  temperatures at 3$\sigma$ confidence for \rratc\ (thick blue lines)
  at an assumed distance of 3.4\,kpc, and \rratb\ (thin red lines) at
  an assumed distance of 5.2\,kpc, given the non-detection of X-ray
  sources in our \chandra\ images.  We assume a radius of 10\,km
  (dashed lines) or 20\,km (solid lines).  We take a range of interstellar
  absorption \NH, where the arrows show the estimates of \NH\ from
  various methods as labelled: the extinction model of \citet{dcllc03}
  as ``$A_V$'', the RRATs dispersion measure as ``DM'', Galactic {\sc
  H\,I} data of \citet{dl90} as ``{\sc H\,I}'', and the
  PSR~J1846$-$0258/Kes~75 complex as ``SNR.''  Values for \rrata\ are
  on the top, values for \rratb\ are on the bottom, and the
  intersections with the luminosity limits are marked with circles and
  squares, respectively.  The dotted line shows the effective
  temperature of \rrata.  
\textit{Right:} Upper limits to the unabsorbed X-ray luminosities at
  3$\sigma$ confidence in the 0.3--8\,keV range for \rratc\ (thick blue
  lines) at an assumed distance of 3.4\,kpc, and \rratb\ (thin red
  lines) at an assumed distance of 5.2\,kpc.  We assume a blackbody
  model with temperature of 70\,eV (solid line), 140\,eV (dashed
  line), or 210\,keV (dot-dashed line) for a range of interstellar
  absorption \NH.  The dotted line shows the luminosity of \rrata\
  ($\expnt{4}{33}\,\ergsec$).  The arrows and filled symbols are the
  same as in the left panel.}
\label{fig:thermal}
\end{figure*}

We determine luminosity limits in two ways.  First, we assume that the
entire surface of the RRATs emits as a blackbody, and we use our
count-rate limits to set upper limits to the surface temperature
(measured at infinity, like all blackbody parameters discussed below).
Blackbody emission is not necessarily the best assumption as many
other atmosphere models are more realistic (e.g., \citealt{rbg+06}
found hydrogen atmospheres gave acceptable fits for \rrata), but we
keep it for simplicity.  In Figure~\ref{fig:thermal} we give
temperature limits both for a canonical 10\,km neutron star radius and
a 20\,km radius more similar to that inferred for \rrata\
\citep{rbg+06}, covering the wide range in \NH\ possible for each
source.  [The larger radius is not likely physical, and may reflect a
combination of errors in the distance or unrealistic models for the
surface emission.  However, more realistic models (.e.g, model
atmospheres) typically have \textit{larger} areas than blackbodies.]

As an alternate method, since assuming a constant emission area is not
always consistent with observations of strongly pulsed neutron stars,
we take the thermal spectrum measured for \rrata: a blackbody with
$kT=140\,$eV and unabsorbed luminosity $\expnt{4}{33}\,\ergsec$
\citep{mrg+07}.  Again considering the wide range in \NH\ possible for
each source, we plot the limits on the unabsorbed thermal luminosity
in Figure~\ref{fig:thermal} for three blackbody temperatures: the
nominal value 140\,eV, and $\pm50$ per cent of that value.  The main
goal of the range in temperature is to show the effects of
interstellar absorption and the limited energy window of \chandra\ on
the luminosity limits, although we physically motivate them in
Section~\ref{sec:disc}.

Assuming blackbody emission from a 20\,km radius and the nominal
absorption, the effective temperature limit on \rratc\ is 77\,eV
(using $\NH=\expnt{5}{21}\,\percmsq$).  While the count-rate limit is
tighter for \rratb, the higher likely distance and absorption
(although the DM is lower) mean that the limit is comparable, 91\,eV
(for $\NH=\expnt{2}{22}\,\percmsq$).  Alternately, if we take
blackbody emission at a fixed temperature of 140\,eV we get luminosity
limits of $\expnt{1}{32}d_{3.4}^2\,\ergsec$ and
$\expnt{3}{32}d_{5.2}^2\,\ergsec$.  
The temperature limits for both sources are considerably cooler than
for \rrata\ for almost all absorptions values.  Similarly, The
luminosity limits for both sources are typically below that of \rrata\
for a range of input spectra, often by several orders of magnitude.
Only for the lowest temperature considered (70\,eV) and absorptions at
the high end of the considered range do our limits become less
constraining, although we must consider the correlation between
distance and \NH, where a larger true distance would also be
associated with more absorption and hence an even weaker limit.  We
also consider a non-thermal spectrum (power-law with photon index of
2), which gives limits of $\expnt{4}{31}d_{3.4}^2\,\ergsec$ and
$\expnt{8}{30}d_{5.2}^2\,\ergsec$ for \RRATc\ and \RRATb.

\section{ Discussion and Conclusions}
\label{sec:disc}
The X-ray emission from \rrata\ was assumed to be largely due to
cooling emission from the neutron star surface \citep{rbg+06,mrg+07},
with only a small contribution from an extended nebula \citep{rmg+09}.
If we expect similar emission from RRATs~\RRATc\ and \RRATb, it would
be diminished due to their older ages (characteristic ages of 0.8 and
0.4\,Myr respectively, vs.\ 0.1\,Myr for \rrata).  In the
neutrino-dominated cooling regime, the temperature declines slowly
with time, with surface temperature $\sim t^{-1/12}$ \citep{plps04}.
The transition to photon-dominated cooling typically occurs around
0.1--1.0\,Myr, and after that the decline is much steeper and
dependent on the nature of the envelope, but exponents of 1--3 are
common.  So in the worst case, and assuming that characteristic age is
correlated with true age (something that is not necessarily true;
e.g., \citealt{gf00,klh+03,kvk09}) we would expect tiny temperatures
$<10\,$eV for RRATs~\RRATc\ and \RRATb\ that would be undetectable.
However, we know of other neutron stars with similar characteristic
ages with luminosities of $\sim 10^{33}\,\ergsec$ (e.g.,
PSR~B1055$-$52 with characteristic age 0.5\,Myr; \citealt{dlcm+05}).
We also know that characteristic age (and even true age) do not always
correlate strictly with effective temperature \citep{vkk08}.  So while
it is tempting to say that the RRATs studied here are older and colder
than \rrata, that may be misleading.  While the unknown distance and
column densities limit the strength of any conclusions, our data may
actually provide upper limits on the luminosities of two sources that
point to a range in X-ray emission among the RRATs.

The RRATs lie close to the INS in the $P$-$\dot P$ plane in a region
with few other pulsars, and appear like the INS to have preferentially
long periods (although there are substantial selection effects;
\citealt{dcm+09,mlk+09}).  Our X-ray non-detections of the RRATs are
actually consistent over most of the range in $\NH$ with emission like
that seen from the INS: blackbodies with $kT=40-100\,$eV and
luminosities $\sim 10^{32}\,\ergsec$.  They are generally consistent
with a cooling sequence that also includes the younger pulsars like
PSR~B0656+14 and PSR~B1055$-$52 \citep{plps09}, although the details
of the emission are difficult and there could be a small contribution
from magnetic field decay \citep[][and see below]{kvk09}.  While the
characteristic ages of the INS are a factor of 4--8 older than those
of the RRATs considered here, the true ages are likely comparable
(\citealt{msh+05,mph+09}; \citealt*{kvka02,kvka07}).  If the RRATs and
the INS were drawn from a single cooling sequence that evolved with
the characteristic age, we would expect \RRATc\ and \RRATb\ to lie
somewhere between 140\,eV (\RRATa) and 70\,eV (the mean of the INS),
with the luminosity declining accordingly.  In that case the RRATs
could be at $\sim 10^{32}\,\ergsec$ for much of the considered range
in $\NH$, and significantly deeper observations would be required to
understand the true nature of their emission.  Considering true age
rather than characteristic age complicates the situation as the RRATs
do not have any such measurements, but the general argument still
holds.

The X-ray luminosity of \rrata\ is above the spin-down luminosity
$\dot E$ ($\sim \expnt{3}{32}\,\ergsec$), implying that
rotationally-powered non-thermal
processes cannot drive the X-ray emission, contrary to what is seen
for many rotation-powered pulsars \citep[$L_X\sim 10^{-3}\dot
E$][]{bt97,pccm02,ctw06} but consistent with the INS. The non-thermal
contributions to the X-ray emission from RRATs~\RRATc\ and \RRATb\
would be $10^{28-29}\,\ergsec$, below our limits even for optimistic
values of the distance and \NH.  Therefore our limits cannot really
constrain the expected level of non-thermal emission from these
sources.  We note, though, that the observed spin-down luminosity
appears too low to power the extended X-ray emission observed around
\rrata\ (interpreting it as a pulsar wind nebula) requiring some other
energy source \citep{rmg+09}, which is also a possible conclusion
about the nebula around the INS RX~J1856.5$-$3754 \citep{vkk08}.

A final possibility for the emission of \rrata\ is that, while the
spectrum is thermal, the emission we see is cooling augmented by
magnetic field decay \citep[e.g.,][and references
therein]{hk98,pmg09}.  While the exact field decay mechanisms and
timescales are still uncertain \citep{gr92}, field decay may be
relevant for objects of ages $<1\,$Myr and magnetic fields
$>\expnt{2}{13}\,$G  \citep{pmg09}, which is the correct range for the
RRATs considered here.
We would expect generally that for objects of similar ages (such that
the same decay mechanisms are operating) the luminosity due to field
decay would scale as $B |\dot B|$ \citep{hk98}, where $|\dot B|\sim
B/\tau_B$ and $\tau_B$ is the relevant decay timescale which may
itself depend on $B$ \citep{hh97}.  There are many complexities and
dependencies (magnetic field configuration, temperature effects, etc.)
that we must ignore here, but if the heating is due to Ohmic diffusion
\citep{pmg09} then the timescale is independent of $B$ and the
luminosity should scale as $B^2$.  So the factor of $\sim 2$
difference in dipolar magnetic field between \rrata\ and the others
would lead to at most a factor of 4 difference in instantaneous
thermal luminosity.  Once again, this is consistent with our limits
but the large uncertainties in the distances and absorptions make
conclusions difficult.  Additionally, this energy takes time to diffuse to
the surface, so the flux we see now may represent a different set of
conditions in the past when the difference in magnetic field was even
wider, if decay makes the field strength  converge to a single value
as discussed by \citet{pmg09}.


In summary, deep \chandra\ searches at the positions of RRATs \RRATc\
and \RRATb\ show no X-ray emission from these sources.  In order to
derive limits on the X-ray luminosity, we have considered several
models for the emission based on the one RRAT with an X-ray
counterpart (\RRATa).  The simplest interpretation is thermal emission
from a cooling neutron star, but non-thermal emission driven by the
magnetosphere and thermal emission driven by magnetic field decay are
also possibilities.  Blackbody emission from a 20\,km radius (like
\rrata) leads to temperature limits of 77 and 91\,eV.  Assuming
thermal emission similar to that seen from \rrata\ (a blackbody with
$kT=0.14\,$keV), we derive luminosity limits in the 0.3--8\,keV range
of $\expnt{1}{32}d_{3.4}^2\,\ergsec$ and
$\expnt{3}{32}d_{5.2}^2\,\ergsec$.  However, with only a single source
to use as a template, \rrata, it is unclear if our non-detections come
from innate differences in the sources (the sources we studied are
likely slightly older and have lower magnetic fields), an
inhomogeneous population, or errors in the distances and column
densities of any of the sources.

\section*{Acknowledgments}
We thank an anonymous referee for helpful comments.  DLK was supported
by NASA through Hubble Fellowship grant \#01207.01-A awarded by the
Space Telescope Science Institute, which is operated by the
Association of Universities for Research in Astronomy, Inc., for NASA,
under contract NAS 5-26555.  We acknowledge support through
\textit{Chandra} grant GOO7-8064X.  POS acknowledges partial support
from NASA Contract NAS8-03060.  This research has made use of software
provided by the Chandra X-ray Center (CXC) in the application package
CIAO.


\begin{thebibliography}{}

\bibitem[\protect\citeauthoryear{{Becker} \& {Tr\"{u}mper}}{{Becker} \&
  {Tr\"{u}mper}}{1997}]{bt97}
{Becker} W.,  {Tr\"{u}mper} J.,  1997, \aap, 326, 682

\bibitem[\protect\citeauthoryear{{Brown}}{{Brown}}{1995}]{brown95}
{Brown} G.~E.,  1995, \apj, 440, 270

\bibitem[\protect\citeauthoryear{{Cheng}, {Taam} \& {Wang}}{{Cheng}
  et~al.}{2006}]{ctw06}
{Cheng} K.~S.,  {Taam} R.~E.,    {Wang} W.,  2006, \apj, 641, 427

\bibitem[\protect\citeauthoryear{Cordes \& Lazio}{Cordes \& Lazio}{2002}]{cl02}
Cordes J.~M.,  Lazio T. J.~W.,  2002, astro-ph/0207156

\bibitem[\protect\citeauthoryear{{Cordes}, {Lazio} \& {McLaughlin}}{{Cordes}
  et~al.}{2004}]{clm04}
{Cordes} J.~M.,  {Lazio} T.~J.~W.,    {McLaughlin} M.~A.,  2004, New Astronomy
  Review, 48, 1459, (astro-ph/0410045)

\bibitem[\protect\citeauthoryear{{Cordes} \& {Shannon}}{{Cordes} \&
  {Shannon}}{2008}]{cs08}
{Cordes} J.~M.,  {Shannon} R.~M.,  2008, \apj, 682, 1152

\bibitem[\protect\citeauthoryear{{De Luca}, {Caraveo}, {Mereghetti}, {Negroni}
  \& {Bignami}}{{De Luca} et~al.}{2005}]{dlcm+05}
{De Luca} A.,  {Caraveo} P.~A.,  {Mereghetti} S.,  {Negroni} M.,    {Bignami}
  G.~F.,  2005, \apj, 623, 1051

\bibitem[\protect\citeauthoryear{{Deneva} et~al.,}{{Deneva}
  et~al.}{2009}]{dcm+09}
{Deneva} J.~S.,  et~al., 2009, \apj, submitted

\bibitem[\protect\citeauthoryear{{Dickey} \& {Lockman}}{{Dickey} \&
  {Lockman}}{1990}]{dl90}
{Dickey} J.~M.,  {Lockman} F.~J.,  1990, \araa, 28, 215

\bibitem[\protect\citeauthoryear{{Drimmel}, {Cabrera-Lavers} \& {L{\'
  o}pez-Corredoira}}{{Drimmel} et~al.}{2003}]{dcllc03}
{Drimmel} R.,  {Cabrera-Lavers} A.,    {L{\' o}pez-Corredoira} M.,  2003, \aap,
  409, 205

\bibitem[\protect\citeauthoryear{{Eikenberry}, {Matthews}, {LaVine}, {Garske},
  {Hu}, {Jackson}, {Patel}, {Barry}, {Colonno}, {Houck}, {Wilson}, {Corbel} \&
  {Smith}}{{Eikenberry} et~al.}{2004}]{eml+04}
{Eikenberry} S.~S.,  {et al.}  2004, \apj, 616, 506

\bibitem[\protect\citeauthoryear{{Gaensler} \& {Frail}}{{Gaensler} \&
  {Frail}}{2000}]{gf00}
{Gaensler} B.~M.,  {Frail} D.~A.,  2000, \nat, 406, 158

\bibitem[\protect\citeauthoryear{{Gaensler}, {McClure-Griffiths}, {Oey},
  {Haverkorn}, {Dickey} \& {Green}}{{Gaensler} et~al.}{2005}]{gmgo+05}
{Gaensler} B.~M.,  {McClure-Griffiths} N.~M.,  {Oey} M.~S.,  {Haverkorn} M.,
  {Dickey} J.~M.,    {Green} A.~J.,  2005, \apjl, 620, L95

\bibitem[\protect\citeauthoryear{{Garmire}, {Bautz}, {Ford}, {Nousek} \&
  {Ricker}}{{Garmire} et~al.}{2003}]{gbf+03}
{Garmire} G.~P.,  {Bautz} M.~W.,  {Ford} P.~G.,  {Nousek} J.~A.,    {Ricker}
  G.~R.,  2003, \procspie, 4851, 28

\bibitem[\protect\citeauthoryear{{Gavriil}, {Gonzalez}, {Gotthelf}, {Kaspi},
  {Livingstone} \& {Woods}}{{Gavriil} et~al.}{2008}]{ggg+08}
{Gavriil} F.~P.,  {Gonzalez} M.~E.,  {Gotthelf} E.~V.,  {Kaspi} V.~M.,
  {Livingstone} M.~A.,    {Woods} P.~M.,  2008, Science, 319, 1802

\bibitem[\protect\citeauthoryear{{Gehrels}}{{Gehrels}}{1986}]{gehrels86}
{Gehrels} N.,  1986, \apj, 303, 336

\bibitem[\protect\citeauthoryear{{Goldreich} \& {Reisenegger}}{{Goldreich} \&
  {Reisenegger}}{1992}]{gr92}
{Goldreich} P.,  {Reisenegger} A.,  1992, \apj, 395, 250

\bibitem[\protect\citeauthoryear{{Gotthelf} \& {Halpern}}{{Gotthelf} \&
  {Halpern}}{2005}]{gh05}
{Gotthelf} E.~V.,  {Halpern} J.~P.,  2005, \apj, 632, 1075

\bibitem[\protect\citeauthoryear{{Haberl}}{{Haberl}}{2007}]{haberl07}
{Haberl} F.,  2007, \apss, 308, 181

\bibitem[\protect\citeauthoryear{{Helfand}, {Collins} \& {Gotthelf}}{{Helfand}
  et~al.}{2003}]{hcg03}
{Helfand} D.~J.,  {Collins} B.~F.,    {Gotthelf} E.~V.,  2003, \apj, 582, 783

\bibitem[\protect\citeauthoryear{{Heyl} \& {Hernquist}}{{Heyl} \&
  {Hernquist}}{1997}]{hh97}
{Heyl} J.~S.,  {Hernquist} L.,  1997, \apjl, 491, L95

\bibitem[\protect\citeauthoryear{{Heyl} \& {Kulkarni}}{{Heyl} \&
  {Kulkarni}}{1998}]{hk98}
{Heyl} J.~S.,  {Kulkarni} S.~R.,  1998, \apjl, 506, L61

\bibitem[\protect\citeauthoryear{{Hyman}, {Lazio}, {Kassim}, {Ray}, {Markwardt}
  \& {Yusef-Zadeh}}{{Hyman} et~al.}{2005}]{hlk+05}
{Hyman} S.~D.,  {Lazio} T.~J.~W.,  {Kassim} N.~E.,  {Ray} P.~S.,  {Markwardt}
  C.~B.,    {Yusef-Zadeh} F.,  2005, \nat, 434, 50

\bibitem[\protect\citeauthoryear{{Ibrahim}, {Markwardt}, {Swank}, {Ransom},
  {Roberts}, {Kaspi}, {Woods}, {Safi-Harb}, {Balman}, {Parke}, {Kouveliotou},
  {Hurley} \& {Cline}}{{Ibrahim} et~al.}{2004}]{ims+04}
{Ibrahim} A.~I.,  {et al.}  2004, \apjl, 609, L21

\bibitem[\protect\citeauthoryear{{Kaplan}}{{Kaplan}}{2008}]{kaplan08}
{Kaplan} D.~L.,  2008, AIPC, 983, 331, arXiv:0801.1143

\bibitem[\protect\citeauthoryear{{Kaplan} \& {van Kerkwijk}}{{Kaplan} \& {van
  Kerkwijk}}{2009}]{kvk09}
{Kaplan} D.~L.,  {van Kerkwijk} M.~H.,  2009, \apjl, 692, L62

\bibitem[\protect\citeauthoryear{{Kaplan}, {van Kerkwijk} \&
  {Anderson}}{{Kaplan} et~al.}{2002}]{kvka02}
{Kaplan} D.~L.,  {van Kerkwijk} M.~H.,    {Anderson} J.,  2002, \apj, 571, 447

\bibitem[\protect\citeauthoryear{{Kaplan}, {van Kerkwijk} \&
  {Anderson}}{{Kaplan} et~al.}{2007}]{kvka07}
{Kaplan} D.~L.,  {van Kerkwijk} M.~H.,    {Anderson} J.,  2007, \apj, 660, 1428

\bibitem[\protect\citeauthoryear{{Keane} \& {Kramer}}{{Keane} \&
  {Kramer}}{2008}]{kk08}
{Keane} E.~F.,  {Kramer} M.,  2008, \mnras, 391, 2009

\bibitem[\protect\citeauthoryear{{Klose}, {Henden}, {Geppert}, {Greiner},
  {Guetter}, {Hartmann}, {Kouveliotou}, {Luginbuhl}, {Stecklum} \&
  {Vrba}}{{Klose} et~al.}{2004}]{khg+04}
{Klose} S.,  {et al.}  2004, \apjl, 609, L13

\bibitem[\protect\citeauthoryear{{Knight}, {Bailes}, {Manchester}, {Ord} \&
  {Jacoby}}{{Knight} et~al.}{2006}]{kbm+06}
{Knight} H.~S.,  {Bailes} M.,  {Manchester} R.~N.,  {Ord} S.~M.,    {Jacoby}
  B.~A.,  2006, \apj, 640, 941

\bibitem[\protect\citeauthoryear{{Koptsevich}, {Pavlov}, {Zharikov}, {Sokolov},
  {Shibanov} \& {Kurt}}{{Koptsevich} et~al.}{2001}]{kpz+01}
{Koptsevich} A.~B.,  {Pavlov} G.~G.,  {Zharikov} S.~V.,  {Sokolov} V.~V.,
  {Shibanov} Y.~A.,    {Kurt} V.~G.,  2001, \aap, 370, 1004

\bibitem[\protect\citeauthoryear{{Kramer}, {Lyne}, {Hobbs}, {L{\" o}hmer},
  {Carr}, {Jordan} \& {Wolszczan}}{{Kramer} et~al.}{2003}]{klh+03}
{Kramer} M.,  {Lyne} A.~G.,  {Hobbs} G.,  {L{\" o}hmer} O.,  {Carr} P.,
  {Jordan} C.,    {Wolszczan} A.,  2003, \apjl, 593, L31

\bibitem[\protect\citeauthoryear{{Leahy} \& {Tian}}{{Leahy} \&
  {Tian}}{2008}]{lt08}
{Leahy} D.~A.,  {Tian} W.~W.,  2008, \aap, 480, L25

\bibitem[\protect\citeauthoryear{{Li}}{{Li}}{2006}]{li06}
{Li} X.-D.,  2006, \apjl, 646, L139

\bibitem[\protect\citeauthoryear{{Lorimer} \& {Kramer}}{{Lorimer} \&
  {Kramer}}{2004}]{lk04}
{Lorimer} D.~R.,  {Kramer} M.,  2004, {Handbook of Pulsar Astronomy}.
Cambridge University Press, Cambridge, UK

\bibitem[\protect\citeauthoryear{{Luo} \& {Melrose}}{{Luo} \&
  {Melrose}}{2007}]{lm07}
{Luo} Q.,  {Melrose} D.,  2007, \mnras, 378, 1481

\bibitem[\protect\citeauthoryear{{McLaughlin} et~al.,}{{McLaughlin}
  et~al.}{2006}]{mll+06}
{McLaughlin} M.~A.,  et~al., 2006, \nat, 439, 817

\bibitem[\protect\citeauthoryear{{McLaughlin}, {Lyne}, {Keane}, {Kramer},
  {Miller}, {Lorimer} \& {Manchester}}{{McLaughlin} et~al.}{2009}]{mlk+09}
{McLaughlin} M.~A.,  {Lyne} A.~G.,  {Keane} E.,  {Kramer} M.,  {Miller} J.,
  {Lorimer} D.~R.,    {Manchester} R.~N.,  2009, \mnras, submitted

\bibitem[\protect\citeauthoryear{{McLaughlin}, {Rea}, {Gaensler}, {Chatterjee},
  {Camilo}, {Kramer}, {Lorimer}, {Lyne}, {Israel} \& {Possenti}}{{McLaughlin}
  et~al.}{2007}]{mrg+07}
{McLaughlin} M.~A.,  {et al.}  2007, \apj, 670, 1307

\bibitem[\protect\citeauthoryear{{Motch} et~al.,}{{Motch}
  et~al.}{2005}]{msh+05}
{Motch} C.,  et~al., 2005, \aap, 429, 257

\bibitem[\protect\citeauthoryear{{Motch}, {Pires}, {Haberl}, {Schwope} \&
  {Zavlin}}{{Motch} et~al.}{2009}]{mph+09}
{Motch} C.,  {Pires} A.~M.,  {Haberl} F.,  {Schwope} A.,    {Zavlin} V.~E.,
  2009, \aap, 497, 423

\bibitem[\protect\citeauthoryear{{Muno}, {Clark}, {Crowther}, {Dougherty}, {de
  Grijs}, {Law}, {McMillan}, {Morris}, {Negueruela}, {Pooley}, {Portegies
  Zwart} \& {Yusef-Zadeh}}{{Muno} et~al.}{2006}]{mcc+06}
{Muno} M.~P.,  {et al.}  2006, \apjl,
  636, L41

\bibitem[\protect\citeauthoryear{{Ng}, {Slane}, {Gaensler} \& {Hughes}}{{Ng}
  et~al.}{2008}]{nsgh08}
{Ng} C.-Y.,  {Slane} P.~O.,  {Gaensler} B.~M.,    {Hughes} J.~P.,  2008, \apj,
  686, 508

\bibitem[\protect\citeauthoryear{{Page}, {Lattimer}, {Prakash} \&
  {Steiner}}{{Page} et~al.}{2004}]{plps04}
{Page} D.,  {Lattimer} J.~M.,  {Prakash} M.,    {Steiner} A.~W.,  2004, \apjs,
  155, 623

\bibitem[\protect\citeauthoryear{{Page}, {Lattimer}, {Prakash} \&
  {Steiner}}{{Page} et~al.}{2009}]{plps09}
{Page} D.,  {Lattimer} J.~M.,  {Prakash} M.,    {Steiner} A.~W.,  2009, \apj,
  submitted, arXiv:0906.1621

\bibitem[\protect\citeauthoryear{{Pons}, {Miralles} \& {Geppert}}{{Pons}
  et~al.}{2009}]{pmg09}
{Pons} J.~A.,  {Miralles} J.~A.,    {Geppert} U.,  2009, \aap, 496, 207

\bibitem[\protect\citeauthoryear{{Popov}, {Turolla} \& {Possenti}}{{Popov}
  et~al.}{2006}]{ptp06}
{Popov} S.~B.,  {Turolla} R.,    {Possenti} A.,  2006, \mnras, 369, L23

\bibitem[\protect\citeauthoryear{{Possenti}, {Cerutti}, {Colpi} \&
  {Mereghetti}}{{Possenti} et~al.}{2002}]{pccm02}
{Possenti} A.,  {Cerutti} R.,  {Colpi} M.,    {Mereghetti} S.,  2002, \aap,
  387, 993

\bibitem[\protect\citeauthoryear{{Predehl} \& {Schmitt}}{{Predehl} \&
  {Schmitt}}{1995}]{ps95}
{Predehl} P.,  {Schmitt} J.~H.~M.~M.,  1995, \aap, 293, 889

\bibitem[\protect\citeauthoryear{{Rea}, {McLaughlin}, {Gaensler}, {Slane},
  {Stella}, {Reynolds}, {Burgay}, {Israel}, {Possenti} \& {Chatterjee}}{{Rea}
  et~al.}{2009}]{rmg+09}
{Rea} N.,  {et al.}  2009, \apj, in press, arXiv:0906.1394

\bibitem[\protect\citeauthoryear{{Reynolds}, {Borkowski}, {Gaensler}, {Rea},
  {McLaughlin}, {Possenti}, {Israel}, {Burgay}, {Camilo}, {Chatterjee},
  {Kramer}, {Lyne} \& {Stairs}}{{Reynolds} et~al.}{2006}]{rbg+06}
{Reynolds} S.~P.,  {et al.}  2006, \apjl, 639,
  L71

\bibitem[\protect\citeauthoryear{{van Kerkwijk} \& {Kaplan}}{{van Kerkwijk} \&
  {Kaplan}}{2007}]{vkk07}
{van Kerkwijk} M.~H.,  {Kaplan} D.~L.,  2007, \apss, 308, 191

\bibitem[\protect\citeauthoryear{{van Kerkwijk} \& {Kaplan}}{{van Kerkwijk} \&
  {Kaplan}}{2008}]{vkk08}
{van Kerkwijk} M.~H.,  {Kaplan} D.~L.,  2008, \apjl, 673, L163

\bibitem[\protect\citeauthoryear{{Vlemmings}, {Cordes} \&
  {Chatterjee}}{{Vlemmings} et~al.}{2004}]{vcc04}
{Vlemmings} W.~H.~T.,  {Cordes} J.~M.,    {Chatterjee} S.,  2004, \apj, 610,
  402

\bibitem[\protect\citeauthoryear{{Weisskopf}, {Tananbaum}, {Van Speybroeck} \&
  {O'Dell}}{{Weisskopf} et~al.}{2000}]{wtvso00}
{Weisskopf} M.~C.,  {Tananbaum} H.~D.,  {Van Speybroeck} L.~P.,    {O'Dell}
  S.~L.,  2000, \procspie, 4012, 2

\bibitem[\protect\citeauthoryear{{Weltevrede}, {Stappers}, {Rankin} \&
  {Wright}}{{Weltevrede} et~al.}{2006}]{wsrw06}
{Weltevrede} P.,  {Stappers} B.~W.,  {Rankin} J.~M.,    {Wright} G.~A.~E.,
  2006, \apjl, 645, L149

\bibitem[\protect\citeauthoryear{{Woods} \& {Thompson}}{{Woods} \&
  {Thompson}}{2006}]{wt06}
{Woods} P.~M.,  {Thompson} C.,  2006, in {Lewin} W.,  {van der Klis} M.,  eds,
  Compact stellar X-ray sources {Soft gamma repeaters and anomalous X-ray
  pulsars: magnetar candidates}.
Cambridge University Press, Cambridge, UK, p.~547

\bibitem[\protect\citeauthoryear{{Zhang}, {Gil} \& {Dyks}}{{Zhang}
  et~al.}{2007}]{zgd07}
{Zhang} B.,  {Gil} J.,    {Dyks} J.,  2007, \mnras, 374, 1103

\end{thebibliography}


\end{document}